\documentclass[10pt,conference]{IEEEtran}

\usepackage{epsfig,graphicx,subfigure,psfrag,amsmath,cases,bm}
\usepackage{latexsym,amssymb,algorithm,mathtools}
\usepackage{algorithmic}
\usepackage{color}
\usepackage{url}
\usepackage{scrtime}
\usepackage{stfloats}
\usepackage{tablefootnote}
\usepackage{cite}
\usepackage{amsfonts,mathabx}
\usepackage{textcomp}
\usepackage{xcolor,capt-of}
\usepackage{multirow}
\usepackage[left=0.625in,top=0.73in,bottom=0.95in,right=0.625in]{geometry}

\author{\textit{(Invited Paper)}\vspace*{1mm}\\
\IEEEauthorblockN {Dongfang Xu\IEEEauthorrefmark{1}, Chang Liu\IEEEauthorrefmark{2}, Shenghui Song\IEEEauthorrefmark{1}, and Derrick Wing Kwan Ng\IEEEauthorrefmark{3}\\}
\IEEEauthorrefmark{1}The Hong Kong University of Science and Technology, Hong Kong;\\ \IEEEauthorrefmark{2}The Hong Kong Polytechnic University, Hong Kong;
\IEEEauthorrefmark{3}The University
of New South Wales, Australia

}

\newtheorem{T-Prob}{Transformed Problem}

\DeclareMathOperator{\mino}{minimize}

\allowdisplaybreaks

\makeatletter
\def\ps@IEEEtitlepagestyle{%
  \def\@oddfoot{\mycopyrightnotice}%
  \def\@evenfoot{}%
}
\def\mycopyrightnotice{%
  {\footnotesize 978-1-6654-5245-8/23/31.00 ©2023 IEEE\hfill}
  \gdef\mycopyrightnotice{}
}
\title{Integrated Sensing and Communication in Coordinated Cellular Networks}

\begin{document}
\maketitle
\begin{abstract}
Integrated sensing and communication (ISAC) is a promising technique to provide sensing services in future wireless networks. Numerous existing works have adopted a monostatic radar architecture to realize ISAC, i.e., employing the same base station (BS) to transmit the ISAC signal and receive the echo. Yet, the concurrent information transmission causes unavoidable self-interference (SI) to the radar echo at the BS. To overcome this difficulty, we propose a coordinated cellular network-supported multistatic radar architecture to implement ISAC, which allows us to spatially separate the ISAC signal transmission and radar echo reception, intrinsically circumventing the problem of SI. To this end, we jointly optimize the transmit and receive beamforming policy to minimize the sensing beam pattern mismatch error subject to ISAC quality-of-service requirements. The resulting non-convex optimization problem is tackled by an alternating optimization-based suboptimal algorithm. Simulation results showed that the proposed scheme outperforms the two baseline schemes adopting conventional designs.
\end{abstract}

\section{Introduction}
Recently, integrated sensing and communication (ISAC) has drawn considerable research interest due to its appealing advantages, such as enabling efficient spectrum sharing between communication and radar systems, and acquiring sensing information for a variety of innovative applications \cite{9705498}. Motivated by these advantages, various works have proposed to unleash the potential of ISAC to improve the performance of wireless networks, e.g., \cite{8386661,xu2022robust,liu2022learning}. In particular, the authors of \cite{8386661} developed several optimization-based waveform design schemes for minimizing the downlink multiuser interference in a dual-functional radar and communication (DFRC) base stations (BSs)-supported ISAC system. Also, in \cite{xu2022robust}, the authors proposed a novel optimization framework with variable-length time slots to flexibly prioritize communication and sensing so as to achieve secure and robust ISAC. In fact, most of the works in the literature, including \cite{8386661} and \cite{xu2022robust}, adopt a monostatic radar architecture to achieve ISAC, i.e., exploiting the same DFRC BS to transmit ISAC signal and receive radar echo. Yet, in practice, a radar echo usually returns to the DFRC BS before the information transmission ends. For instance, for a sensing target that is $300$ meters away from the DFRC BS, the desired echo signal bounces back in only $2$ $\mu$s. While in the Long-Term Evolution (LTE) standard \cite{ghosh2010fundamentals} and the Fifth-Generation New Radio (5G NR) standard \cite{access2020newradio}, the symbol durations are typically around tens of microseconds. As a result, strong self-interference (SI) is expected between the concurrent ISAC signal transmission and radar echo reception. Besides, echo signals typically suffer from severe round-trip path loss \cite{skolnik2008radar}. Following the example above, the considered sensing target causes a free-space round-trip path loss of roughly $180$ dB for a sensing signal at $2.4$ GHz, which makes the echo signal significantly weaker than the SI. As such, conventional SI cancellation techniques developed for full-duplex communication systems may not be able to effectively suppress the SI at the monostatic BS to below the targeted echo power level \cite{xu2022integrated}. This leads to a bottleneck in realizing high-quality ISAC in wireless networks.
\par
Motivated by the above observations, in this paper, we propose to leverage cellular network-based multistatic radar architecture to achieve high-quality ISAC. Specifically, given a set of coordinated DFRC BSs, we dynamically form a multistatic radar system based on a pre-designed BS selection criterion. The transmit and receive beamforming at the multistatic transmitter and receiver are jointly optimized to minimize the beam pattern mismatch error while ensuring the quality-of-service (QoS) requirements in information transmission and sensing. An alternating optimization (AO)-based algorithm is developed to tackle the formulated optimization problem efficiently.
\par
\textit{Notation:} Vectors and matrices are denoted by boldface lowercase and boldface capital letters, respectively. $\mathbb{R}^{N\times M}$ and $\mathbb{C}^{N\times M}$ denote the space of $N\times M$ real-valued and complex-valued matrices, respectively. $|\cdot|$ and $||\cdot||_2$ denote the absolute value and the $l_2$-norm operators, respectively. $(\cdot)^T$, and $(\cdot)^H$ stand for the transpose and the conjugate transpose of of their arguments, respectively. $\mathbf{I}_{N}$ refers to the identity matrix of dimension $N$. $\mathbb{H}^{N}$ denotes the set of complex Hermitian matrices of dimension $N$. $\mathrm{Tr}(\cdot)$ and $\mathrm{Rank}(\cdot)$ refer to the trace and rank of their arguments, respectively. $\mathrm{Diag}(\mathbf{X})$ represents a vector whose elements are extracted from the main diagonal of matrix $\mathbf{X}$; $\mathrm{diag}(\mathbf{x})$ denotes an $N\times N$ diagonal matrix with main diagonal elements $x_1,\cdots, x_N$. $\mathbf{A}\succeq\mathbf{0}$ indicates that $\mathbf{A}$ is a positive semidefinite matrix. $\mathcal{CN}(0,\sigma^2)$ specifies the distribution of a circularly symmetric complex Gaussian (CSCG) random variable with mean $0$ and variance $\sigma^2$. $\overset{\Delta }{=}$ and $\sim$ stand for ``defined as'' and ``distributed as'', respectively. $\mathcal{E}\left \{ \cdot \right \}$ denotes the statistical expectation. 
\section{System Model}
In this section, we first introduce the proposed multi-cell network-based ISAC system. Subsequently, we present the corresponding signal model.
\begin{figure}[t] 
\centering\includegraphics[width=2.4in]{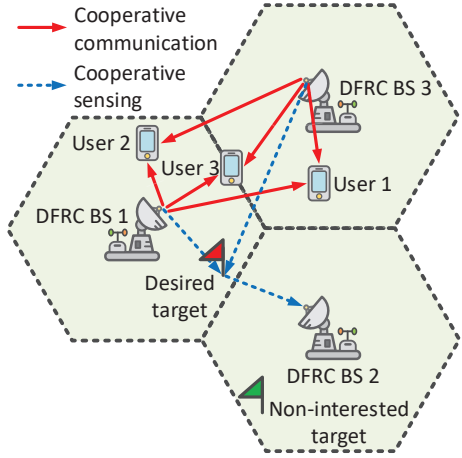}
\caption{Illustration of a cellular network-based integrated sensing and communication system comprising $K=3$ users and $M=3$ coordinated cells, where each cell contains a dual-functional radar and communication (DFRC) base station (BS). Specifically, DFRC BSs $1$ and $3$ cooperate to transmit ISAC signals to serve the $K$ users and sense a desired target, while DFRC BS $2$ is selected as the multistatic receiver to receive echo signal.}\label{fig:1 system}
\end{figure}
\subsection{Coordinated Cellular Network-Based ISAC System Model}
We consider a coordinated multi-cell system comprising $M$ cells, each with a BS located at the center, cf. Fig. \ref{fig:1 system}. In particular, the $M$ BSs synchronize with each other to form a cooperation cluster, i.e., share the spectrum, data, and channel state information (CSI) \cite{bjornson2013optimal}, to provide communication services for $K$ single-antenna users. Also, each BS is assumed to be equipped with a uniform linear array comprising $N$ antenna elements and all the BSs are capable of DFRC to facilitate ISAC. Moreover, there is one desired sensing target and $J$ non-interested targets in the area of the multi-cell network. To effectively suppress the interference from the information transmission to the echo reception, in this paper, we propose to adopt a multistatic radar architecture in the coordinated multi-cell system. In particular, based on the pre-designed criterion\footnote{The selection criterion will be presented in Section III-A.}, we first select a BS from the cooperation BS cluster as a multistatic radar receiver, denoted by BS $M$, while all the other BSs act as multistatic radar transmitters. Also, all the multistatic transmitters apply information beamforming to convey data signals and to synthesize sensing beams to illuminate the desired sensing target. The selected multistatic radar receiver receives the echo signal from the illuminated sensing target for sensing information extraction. Besides, to investigate the performance upper bound of the considered system, we assume that the user's perfect CSI and target's location information are available on each BS \cite{9124713,9838753}. To simplify the notation, we collect the indices of the multistatic transmitters (unselected BSs), users, and targets in sets $\mathcal{M}=\left \{1,\cdots, M-1 \right \}$, $\mathcal{K}=\left \{1,\cdots ,K \right \}$, and $\mathcal{J}=\left \{0,\cdots ,J \right \}$, respectively. Note that index $j=0$ is associated with the desired sensing target.
\subsection{Signal Model}
In each scheduling time slot, BS $m$ generates a signal stream $\mathbf{x}_m=\underset{k\in\mathcal{K}}{\sum}\mathbf{w}_{m,k}b_k$, $\forall m\in\mathcal{M}$. Here, $\mathbf{w}_{m,k}\in\mathbb{C}^{N\times 1}$ and $b_k\in\mathbb{C}$ denote the beamforming vector and the information-carrying symbol dedicated to user $k$, respectively. We assume $\mathcal{E}\{\left |b_k \right|^2\}=1$, $\forall\mathit{k} \in \mathcal{K}$, without loss of generality. The received signals at user $k$ and at the multistatic receiver (BS $M$) are given by, respectively, \cite{9667503}
\begin{eqnarray}
\label{sig_model1}
y_k&\hspace{-2mm}=\hspace{-2mm}&\underset{m\in\mathcal{M}}{\sum}\big(\mathbf{g}_{m,k}^H+\underset{j\in\mathcal{J}}{\sum}l_{k,j}\mathbf{f}^H_{m,j}\big)\mathbf{x}_m+n_k,\\
\label{sig_model2}
\mathbf{y}_M&\hspace{-2mm}=\hspace{-2mm}&\underset{m\in\mathcal{M}}{\sum}\big(\underset{j\in\mathcal{J}}{\sum}\mathbf{h}_{M,j}\mathbf{f}_{m,j}^H+\mathbf{F}_{m,M}\big)\mathbf{x}_m+\mathbf{n}_M.
\end{eqnarray}
Here, $n_k\sim\mathcal{CN}(0,\sigma_k^2)$ and $\mathbf{n}_M\sim\mathcal{CN}(\mathbf{0},\sigma_M^2\mathbf{I}_{N})$ denote the additive white Gaussian noises (AWGNs) at user $k$ with variance $\sigma_k^2$ and at the multistatic receiver with noise variance $\sigma_M^2$, respectively \cite{liu2023predictive}. Also, vector $\mathbf{g}_{m,k}\in \mathbb{C}^{N\times 1}$ and scalar $l_{k,j}\in \mathbb{C}$ characterize the channel between BS $m$, $m\in\mathcal{M}$, and user $k$ and the channel between user $k$ and target $j$, respectively.\footnote{Similar to \cite{9667503}, we assume that the communication users would not cause any signal reflection.} Moreover, matrix $\mathbf{F}_{m,M}\in \mathbb{C}^{N\times N}$ characterizes the channel between BS $m$, $m\in\mathcal{M}$, and the multistatic receiver. Furthermore, vectors $\mathbf{f}_{m,j}\in \mathbb{C}^{N\times 1}$ and $\mathbf{h}_{M,j}\in \mathbb{C}^{N\times 1}$ represent the channel between BS $m$ and target $j$ and the channel between target $j$ and the multistatic receiver, respectively, where $m\in\mathcal{M}$, $j\in\mathcal{J}$. In this paper, we model $\mathbf{f}_{m,j}$ and $\mathbf{h}_{M,j}$ as pure line-of-sight (LoS) channels \cite{8288677}. Specifically, channel vector $\mathbf{f}_{m,j}$ is given by
\begin{equation}
\mathbf{f}_{m,j}=\sqrt{\alpha_{m,j}}\mathbf{a}(\theta_{m,j}),
\end{equation}
where $\alpha_{m,j}\in\mathbb{R}$ and $\mathbf{a}(\theta_{m,j})\in\mathbb{C}^{N\times 1}$ denote the channel gain and the steering vector between BS $m$ and target $j$, respectively. Specifically, scalar parameter $\alpha_{m,j}$ is given by $\alpha_{m,j}=\gamma_j\frac{\mu}{d^2_{m,j}}\in\mathbb{R}$, where $\gamma_j\in\mathbb{R}$ is the radar cross-section of target $j$ \cite{8579200}, and $d_{m,j}$ is the distance between BS $m$ and target $j$. Parameter $\mu$ is given by $\mu=(\frac{c}{4\pi f_c})^2$ and its value depends on the system center frequency $f_c$ and the speed of light $c$. Moreover, vector $\mathbf{a}(\theta_{m,j})$ is given by $\mathbf{a}(\theta_{m,j})=\Big[1,e^{j2\pi\omega\mathrm{sin}\theta_{m,j}},\cdots,e^{j2\pi\omega(N_\mathrm{T}-1)\mathrm{sin}\theta_{m,j}}\Big]^T$, where $\theta_{m,j}$ and $\omega$ are the angle of departure (AoD) from BS $m$ to target $j$ and the normalized spacing between adjacent antenna elements, respectively. Similarly, channel vector $\mathbf{h}_{M,j}$ can be expressed as follows
\begin{equation}
\mathbf{h}_{M,j}=\frac{\sqrt{\gamma_j\mu}}{d_{M,j}}\mathbf{a}(\theta_{M,j}),
\end{equation}
where $d_{M,j}\in\mathbb{R}$ denotes the distance between target $j$ and the multistatic receiver. Also, $\mathbf{a}(\theta_{M,j})\in\mathbb{C}^{N\times 1}$ and $\theta_{M,j}$ are the steering vector between target $j$ and the multistatic receiver and the corresponding angle of arrival (AoA), respectively. For ease of presentation, we rewrite \eqref{sig_model1} and \eqref{sig_model2} equivalently as follows, respectively,
\begin{eqnarray}
\label{sig_model3}
y_k&\hspace{-2mm}=\hspace{-2mm}&\underbrace{\mathbf{g}_k^H\mathbf{w}_kb_k}_{\text{Desired data signal}}+\underbrace{\mathbf{g}_k^H\underset{i\in\mathcal{K}\setminus\left\{k\right\}}{\sum}\mathbf{w}_ib_i}_{\text{Multiuser interference}}+n_k,\\
\label{sig_model4}
\mathbf{y}_M&\hspace{-2mm}=\hspace{-2mm}&\underbrace{\mathbf{h}_{M,0}\mathbf{f}_0^H\underset{k\in\mathcal{K}}{\sum}\mathbf{w}_kb_k}_{\text{Desired echo signal}}+\underbrace{\underset{j\in\mathcal{J}\setminus\left\{0\right\}}{\sum}\mathbf{h}_{M,j}\mathbf{f}_j^H\underset{k\in\mathcal{K}}{\sum}\mathbf{w}_kb_k}_{\text{Undesired clutter}}\notag\\
&\hspace{-2mm}+\hspace{-2mm}&\underbrace{\mathbf{F}_M\underset{k\in\mathcal{K}}{\sum}\mathbf{w}_kb_k}_{\text{BS crosstalk}}+\mathbf{n}_M.
\end{eqnarray}
Here, we define $\mathbf{g}_k\in\mathbb{C}^{N(M-1)\times 1}$, $\mathbf{f}_j\in\mathbb{C}^{N(M-1)\times 1}$, $\mathbf{F}_M\in\mathbb{C}^{N\times N(M-1)}$, and $\mathbf{w}_k\in\mathbb{C}^{N(M-1)\times 1}$ as follows, respectively,
\begin{eqnarray}\hspace{-0.5mm}
\hspace*{-2mm}\mathbf{g}_k&\hspace{-3mm}\overset{\Delta}{=}\hspace{-3mm}&\left [\mathbf{g}_{1,k}^H\hspace{-0.5mm}+\hspace{-1.5mm}\underset{j\in\mathcal{J}}{\sum}l_{k,j}\mathbf{f}^H_{1,j},\cdots,\mathbf{g}_{(M-1),k}^H\hspace{-0.5mm}+\hspace{-1.5mm}\underset{j\in\mathcal{J}}{\sum}l_{k,j}\mathbf{f}^H_{(M-1),j}\right]^H,\\
\hspace*{-2mm}\mathbf{f}_j&\hspace{-3mm}\overset{\Delta}{=}\hspace{-3mm}&\left [\mathbf{f}_{1,j}^H,\cdots,\mathbf{f}_{m,j}^H,\cdots,\mathbf{f}_{(M-1),j}^H\right]^H,\hspace*{1mm}\forall j\in\mathcal{J},\\
\hspace*{-2mm}\mathbf{F}_M&\hspace{-3mm}\overset{\Delta}{=}\hspace{-3mm}&\left [\mathbf{F}_{1,M},\cdots,\mathbf{F}_{m,M},\cdots,\mathbf{F}_{(M-1),M}\right],\\
\hspace*{-2mm}\mathbf{w}_k&\hspace{-3mm}\overset{\Delta}{=}\hspace{-3mm}&\left [\mathbf{w}^H_{1,k},\cdots,\mathbf{w}^H_{m,k},\cdots,\mathbf{w}^H_{(M-1),k}\right]^H,\hspace*{1mm}\forall k.
\end{eqnarray}
\section{Problem Formulation}
In this section, we first propose a multistatic receiver selection criterion. Then, we introduce the adopted QoS metrics and formulate the resource allocation algorithm design as a non-convex problem, which will be handled by developing a computationally-efficient suboptimal algorithm. The optimal algorithm design will be studied in our future work.
\subsection{Distance-Based BS Selection Criterion}
\label{criterion}
First, we propose a low-complexity heuristic multistatic receiver selection criterion based on the geometry of the network as follows
\par
\textit{\underline{Multistatic receiver selection}:}
\begin{eqnarray}
   \hspace*{-2mm}\mbox{\textit{i. Calculation:}}\hspace*{2mm} Q_i&\hspace*{-2mm}=\hspace*{-2mm}&\frac{\rho\left(\underset{k\in\mathcal{K}}{\prod}r_{i,k}^{\beta_{i,k}}\right)^{\frac{1}{K}}}{(1-\rho)d_{i,0}^2},\hspace*{1mm}\forall i\in\left\{1,\cdots,M\right\},\label{TX_metric}\\
   \hspace*{-2mm}\mbox{\textit{ii. Relabel:}}\hspace*{7mm} M&\hspace*{-2mm}=\hspace*{-2mm}&\mathrm{arg}\hspace*{1mm}\underset{\forall i\in\left\{1,\cdots,M\right\}}{\mathrm{max}}\hspace*{2mm}Q_i.\label{TX_selection}
\end{eqnarray}
Here, we calculate $Q_i$ for each BS and relabel the selected BS as BS $M$. Also, the weighted factor $0<\rho<1$ is used to prioritize sensing and communication while a large value of $\rho$ prefers communication over sensing when selecting the multistatic receiver. Moreover, $d_{i,0}\in\mathbb{R}$ denotes the distance between BS $i$ and the desired target. $r_{i,k}\in\mathbb{R}$ and $\beta_{i,k}\in\mathbb{R}$ denote the distance between BS $i$ and user $k$ and the path loss exponent of the corresponding channel, $\forall i\in\left\{1,\cdots,M\right\}$. We interpret the meaning of $Q_i$ as follows. On the one hand, since the multistatic receiver does not contribute to active transmission, we select the farthest BS from the centroid of the region spanned by the locations of the users because the BS's transmission is less power-efficient compared to other BSs; on the other hand, to reduce the path loss attenuation of the echo signal, we prefer to select the BS closest to the sensing target as the multistatic receiver.\footnote{A more comprehensive criterion that also takes into account the non-interested targets will be investigated in our future work.} As a result, $Q_i$ can be regarded as the ratio between the average path loss (in dB) from BS $i$ to all the users and the path loss from BS $i$ to the desired target. By using \eqref{TX_metric} and \eqref{TX_selection}, we select the BS with the maximum $Q_i$ as the multistatic transmitter to facilitate high-quality sensing and communication given limited transmit power. 
\subsection{Performance Metrics}
In each scheduling time slot, the signal-to-interference-plus-noise ratio of user $k$ is given by 
\begin{eqnarray}
\Gamma_k(\mathbf{w}_k)=\frac{\left|\mathbf{g}_k^H\mathbf{w}_k\right|^2}{\underset{i\in\mathcal{K}\setminus  \left\{k\right\}}{\sum}\left|\mathbf{g}_k^H\mathbf{w}_i\right|^2+\sigma_k^2}.
\end{eqnarray}
\par
On the other hand, according to radar sensing theory, the DFRC BS has to illuminate the desired target by transmitting energy-focusing beams with low side lobe leakages such that the desired echoes can be easily distinguished from the non-interested targets-induced clutter \cite{1183861}. To this end, we propose to employ highly-directional beams for sensing. Specifically, we discretize the angular domain $[-\frac{\pi}{2},\frac{\pi}{2}]$ into $I$ directions and define the ideal beam pattern at multistatic transmitter $m$, i.e., $\{P_m(\phi_i)\}_{i=1}^{I}$, $m\in\mathcal{M}$, as \cite{xu2023sensingenhanced}
\begin{equation}
P_m(\phi_i)=\left\{\begin{matrix}
1, & \hspace*{6mm}\left|\phi_i -\theta_{m,0}\right| \leq \frac{\psi}{2} \\
0, & \mbox{otherwise}\\ 
\end{matrix}\right.,\label{idealbeam1}
\end{equation}
where $\psi$ is the desired beamwidth of the ideal beam pattern for sensing the desired target.\footnote{In practice, the value of $\psi$ depends on the location uncertainty of the target, number of antennas, etc.} Since the ideal beam pattern is difficult to generate in practice, we approximate it by properly design the information beamforming vectors. To quantify the accuracy of such an approximation, we adopt the difference between the ideal beam pattern and the actual beam pattern, i.e., the beam pattern mismatch error, as a performance metric for sensing as follows \cite{9838753}
\begin{equation}
C\big(\zeta_m,\mathbf{w}_k\big)\hspace*{-1mm}=\hspace*{-1mm}\sum_{i=1}^{I}\left| \zeta_m P_m(\phi_i)-\mathbf{a}\hspace*{-0.5mm}^H\hspace*{-0.5mm}(\phi_i)\big(\underset{k\in\mathcal{K}}{\sum}\mathbf{w}_k\mathbf{w}_k^H\big)\mathbf{D}_m\mathbf{a}(\phi_i) \right|.
\end{equation}
Here, variable $\zeta_m\in\mathbb{R}$ is used to scale the ideal transmit beam pattern. Also, a diagonal matrix $\mathbf{D}_m\in\mathbb{R}^{N(M-1)\times N(M-1)}$ is defined as $\mathbf{D}_m\overset{\Delta}{=}\mathrm{diag}(\underbrace{0,\cdots,0}_{(m-1)N},\underbrace{1,\cdots,1}_N,\underbrace{0,\cdots,0}_{(M-1-m)N})$, $\forall m\in\mathcal{M}$. Similarly, for the multistatic receiver, we also define the ideal beam pattern $\{P_M(\phi_i)\}_{i=1}^{I}$ and beam pattern mismatch error $E\big(\zeta_M,\mathbf{v}_M\big)$ as follows, respectively,
\begin{eqnarray}
P_M(\phi_i)&\hspace*{-2mm}=\hspace*{-2mm}&\left\{\begin{matrix}
1, & \hspace*{6mm}\left|\phi_i -\theta_{M,0}\right| \leq \frac{\psi}{2} \\
0, & \mbox{otherwise}\\ 
\end{matrix}\right.,\label{idealbeam2}\\
E\big(\zeta_M,\mathbf{v}_M\big)&\hspace*{-2mm}=\hspace*{-2mm}&\sum_{i=1}^{I}\left| \zeta_M P_M(\phi_i)-\mathbf{a}^H(\phi_i)\mathbf{v}_M\mathbf{v}_M^H\mathbf{a}(\phi_i) \right|,
\end{eqnarray}
where variable $\zeta_M\in\mathbb{R}$ is used to scale the ideal receive beam pattern and $\mathbf{v}_M\in\mathbb{C}^{N \times 1}$ is the receive beamforming vector of the multistatic receiver. Moreover, to ensure the reliable detection of the desired echo signal, we consider two additional performance metrics for sensing. On the one hand, to ensure the echo signal of the desired target can be effectively captured, we consider the received power of the desired echo at the multistatic receiver which is given by $P_{\mathrm{S}}(\mathbf{v}_M,\mathbf{w}_k)\overset{\Delta }{=}\mathrm{Tr}(\mathbf{v}_M\mathbf{v}_M^H\mathbf{H}_{0,M}\mathbf{Z}\mathbf{H}_{0,M}^H)$, where $\mathbf{H}_{j,M}\overset{\Delta}{=}\mathbf{h}_{M,j}\mathbf{f}_j^H\in\mathbb{C}^{N \times N(M-1)}$ is the target response matrix of target $j$, $\forall j\in\mathcal{J}$, \cite{8579200}, and $\mathbf{Z}\in\mathbb{C}^{N(M-1) \times N(M-1)}$ is defined as $\mathbf{Z}\overset{\Delta }{=}\underset{k\in\mathcal{K}}{\sum}\mathbf{w}_k\mathbf{w}_k^H$ for notational simplicity. On the other hand, to effectively suppress the interference, we consider the interference at the multistatic receiver, i.e., $I_{\mathrm{S}}(\mathbf{v}_M,\mathbf{w}_k)$, which is given by $I_{\mathrm{S}}(\mathbf{v}_M,\mathbf{w}_k)\overset{\Delta }{=}\mathrm{Tr}\Big(\mathbf{v}_M\mathbf{v}_M^H\big((\underset{j\in\mathcal{J}}{\sum}\mathbf{H}_{j,M}+\mathbf{F}_M)\mathbf{Z}(\underset{j\in\mathcal{J}}{\sum}\mathbf{H}_{j,M}+\mathbf{F}_M)^H
  \mathbf{H}_{0,M}\mathbf{Z}\mathbf{H}_{0,M}^H\big)\Big)$.
\subsection{Optimization Problem Formulation}
In this paper, we aim to minimize the beam pattern mismatch error at both the multistatic transmitters and receiver to facilitate high-quality sensing while satisfying QoS requirements of communication users and target sensing. In particular, the joint transmit and receive beamforming policy, i.e., $\mathbf{w}_k$ and $\mathbf{v}_M$, and scaling factors $\zeta_m$ and $\zeta_M$ can be obtained by solving the following problem
\begin{eqnarray}
\label{prob1}
&&\hspace*{-8mm}\underset{\mathbf{w}_k,\mathbf{v}_M,\zeta_m,\zeta_M}{\mino} \,\, \,\, \underset{m\in\mathcal{M}}{\sum}C\big(\zeta_m,\mathbf{w}_k\big)+E\big(\zeta_M,\mathbf{v}_M\big)\notag\\
&&\hspace*{-4mm}\mbox{s.t.}\hspace*{1mm}\mbox{C1:}\hspace*{1mm}\mathbf{1}_m\mathrm{Diag}(\underset{k\in\mathcal{K}}{\sum}\mathbf{w}_k\mathbf{w}_k^H)\leq P^{\mathrm{max}}_m,\hspace*{1mm}\forall m\in\mathcal{M},\notag\\
&&\hspace*{1mm}\mbox{C2:}\hspace*{1mm}||\mathbf{v}_M||_2^2=1,\hspace*{11mm}\mbox{C3:}\hspace*{1mm}P_{\mathrm{S}}(\mathbf{v}_M,\mathbf{w}_k)\geq P^{\mathrm{req}}_{\mathrm{S}},
\notag\\
&&\hspace*{1mm}\mbox{C4:}\hspace*{1mm}I_{\mathrm{S}}(\mathbf{v}_M,\mathbf{w}_k)\leq I_{\mathrm{S}}^{\mathrm{tol}},\hspace*{1.6mm}\mbox{C5:}\hspace*{1mm}\Gamma_k(\mathbf{w}_k)\geq\Gamma^{\mathrm{req}}_k,\hspace*{0mm}\forall k.
\end{eqnarray}
Here, vector $\mathbf{1}_m\in\mathbb{R}^{1\times (M-1)N}$ is defined as $\mathbf{1}_m\overset{\Delta}{=}[\underbrace{0,\cdots,0}_{(m-1)N},\underbrace{1,\cdots,1}_N,\underbrace{0,\cdots,0}_{(M-1-m)N}]$. In constraint C1, the transmit power of multistatic transmitter $m$ is limited by the power budget $P_m^{\mathrm{max}}>0$. In constraint C2, we take into account the power invariant property in the signal processing at the multistatic receiver. To ensure reliable detection of the echo signal at the multistatic receiver, we restrict the minimum echo power strength and the maximum tolerable interference to be above $P^{\mathrm{req}}_{\mathrm{S}}>0$ and below $I_{\mathrm{S}}^{\mathrm{tol}}>0$, respectively, as specified in constraints C3 and C4, respectively. $\Gamma^{\mathrm{req}}_k>0$ in constraint C5 denotes the minimum required SINR for user $k$. 
\par
Due to the coupled optimization variables and non-convex constraints C3, C4, and C5, \eqref{prob1} is a non-convex problem. In the next section, we develop an AO-based algorithm which produces a high-quality solution to \eqref{prob1} with low computational complexity.
\section{Solution of the Optimization Problem}
To start with, we define beamforming matrices $\mathbf{W}_k=\mathbf{w}_k\mathbf{w}_k^H$, $\forall k\in\mathcal{K}$, and $\mathbf{V}_M=\mathbf{v}_M\mathbf{v}_M^H$. Then, we recast \eqref{prob1} into an equivalent form as follows \cite{wu2023globally}
\begin{eqnarray}
\label{prob2}
&&\hspace{-12mm}\underset{\substack{\mathbf{W}_k\in\mathbb{H}^{N(M-1)},\\\mathbf{W}_k\succeq \mathbf{0},\mathbf{V}_M\in\mathbb{H}^{N},\\\mathbf{V}_M\succeq \mathbf{0},\zeta_m,\zeta_M}}{\mino} \,\,\,\, \underset{m\in\mathcal{M}}{\sum}C\big(\zeta_m,\mathbf{W}_k\big)+E\big(\zeta_M,\mathbf{V}_M\big)\notag\\
&&\hspace*{-12mm}\mbox{s.t.}\hspace*{1mm}\mbox{C1-C5},\mbox{C6:}\hspace*{1mm}\mathrm{Rank}(\mathbf{W}_k)=1,\forall k,\mbox{C7:}\hspace*{0mm}\mathrm{Rank}(\mathbf{V}_M)=1.
\end{eqnarray}
Subsequently, by exploiting AO theory \cite{bezdek2002some}, we divide the feasible set of \eqref{prob1} into two disjoint blocks, i.e., $\left\{\zeta_m,\mathbf{W}_k\right\}$ and $\left\{\zeta_M,\mathbf{V}_M\right\}$, where each block is associated with a subproblem. Each subproblem is solved by fixing the other block. In particular, for given $\left\{\zeta_M,\mathbf{V}_M\right\}$, we can employ semidefinite relaxation (SDR) \cite{luo2010semidefinite,8974403} to remove the rank-one constraint C6. On the other hand, for given $\left\{\zeta_m,\mathbf{W}_k\right\}$, the rank-one constraint C7 is omitted by SDR. The two resulting rank constraint-relaxed subproblems can be optimally solved by applying convex problem solvers such as CVX \cite{grant2008cvx}. The tightness of the SDR is revealed in the following theorem.
\par
\textit{Theorem 1:}\hspace*{1mm}For given $\left\{\zeta_M,\mathbf{V}_M\right\}$, we can always find an optimal beamforming matrix $\mathbf{W}^*_k$ of the rank constraint-relaxed version of the resulting subproblem such that $\mathrm{Rank}(\mathbf{W}^*_k)= 1$. For given $\left\{\zeta_m,\mathbf{W}_k\right\}$, an optimal unit-rank beamforming matrix $\mathbf{V}^*_M$ can always be obtained.
\par
\textit{Proof:} The proof of Theorem 1 follows similar steps as the proof in \cite[Appendix]{9669263}, and is thus omitted here due to page limitation.
\par
We note that for both blocks $\left\{\zeta_m,\mathbf{W}_k\right\}$ and $\left\{\zeta_M,\mathbf{V}_M\right\}$, the associated rank-one relaxed versions of the subproblems are convex problems \cite{yu2020irs,9723093}. As a result, we can optimally solve the two subproblems alternatingly and the objective function value of \eqref{prob1} is monotonically non-increasing. According to \cite{bezdek2002some}, \cite{nocedal1999numerical}, by optimally solving both subproblems, the proposed AO-based iterative algorithm is guaranteed to converge to a stationary point of \eqref{prob1} in polynomial time. Besides, the computational complexity of the AO-based algorithm is given by $\mathcal{O}\Big(\mathrm{log}(1/\tau)\Big(KN^3(M-1)^3+K^2N^2(M-1)^2+N^3\big)\Big)$ \cite{polik2010interior,9183907}, where $\mathcal{O}\left ( \cdot  \right )$ is the big-O notation and $\tau$ is the convergence tolerance of the AO-based algorithm.
\begin{figure}[t] 
\centering\includegraphics[width=3.4in]{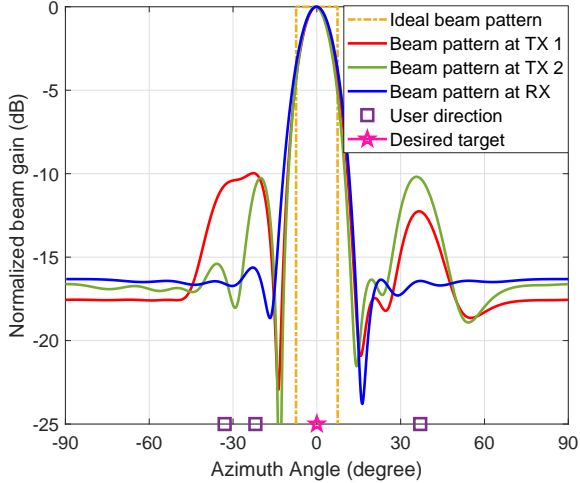}
\caption{Illustration of normalized beam gain of the ideal beam pattern and beam patterns at each BS with $I_{\mathrm{S}}^{\mathrm{tol}}=-96$ dBm.}\label{fig:MulCell_Coop_ISAC_Beampattern}
\end{figure}
\section{Simulation Results}
In this section, we assess the performance of the proposed cooperative ISAC scheme via simulation. In particular, we consider $M=3$ coordinated DFRC BSs and each BS is equipped with $N=9$ antennas.  The $K=3$ users, one desired sensing target, and $J=2$ non-interested target are randomly and uniformly distributed in the multi-cell network and the radius of each cell is $100$ m. For the users-involved channels, i.e., $\mathbf{g}_{m,k}$ and $l_{k,j}$, the path loss exponent is set to $3$, while the path loss exponent of the remaining channels in the system is set to $2$. 
We model the small-scale fading coefficients of the BS-user, BS-BS, and target-user channels as Rician random variables with a Rician factor of $5$. The multistatic receiver is selected based on the criterion developed in Section \ref{criterion}. The angular domain $[-\frac{\pi}{2},\frac{\pi}{2}]$ is equally divided into $I=360$ directions to generate $\{P_m(\phi_i)\}_{i=1}^{I}$, $m\in\mathcal{M}$, and $\{P_M(\phi_i)\}_{i=1}^{I}$. The key parameters are set as $\mu=40$ dB, $\sigma_k^2=-105$ dBm, $P^{\mathrm{req}}_{\mathrm{S}}=-90$ dBm, $\Gamma^{\mathrm{req}}_k=10$ dB, and $P^{\mathrm{max}}_m=43$ dBm \cite{access2020radio}. For comparison, we also consider two baseline schemes. For baseline scheme 1, one DFRC BS employs a monostatic radar architecture for target sensing, while the other two BSs assist in information transmission. For baseline scheme 2, rather than coordinated operations, all three DFRC BSs adopt the monostatic radar architecture to realize ISAC independently. For both baseline schemes, we take into account the interference from ISAC signal transmission to echo signal reception in constraint C4 of \eqref{prob1}. Then, we assume that SI cancellation is applied by the DFRC BS with a SI cancellation coefficient of $-100$ dB \cite{bharadia2013full}.
\par
Fig. \ref{fig:MulCell_Coop_ISAC_Beampattern} illustrates the ideal beam pattern and the actual beam patterns of the multistatic transmitters and receiver for one channel realization. For ease of comparison, we align the beam patterns such that they are centered at $0$-degree. In particular, we observe that the beam pattern of the multistatic receiver largely mimics the main characteristics of the ideal beam pattern. We can also observe that to suppress the interference from the non-interested target, there is a spatial null at around $15$-degree. As for the beam patterns of the multistatic transmitters, apart from the main lobes that closely approximate the ideal beam pattern, there are also several large side lobes. This is because multistatic transmitters need to allocate a portion of the transmit power to satisfy the QoS requirements of the communication users whose locations are different from the desired targets.
\begin{figure}[t] 
\centering\includegraphics[width=3.4in]{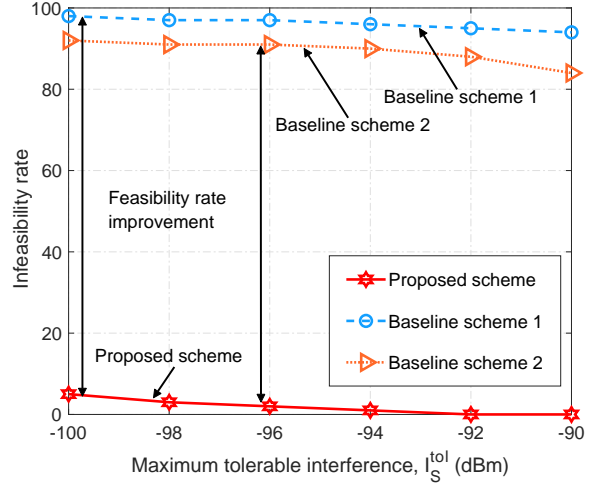}
\caption{Average infeasibility rate versus the maximum tolerable interference for different schemes.}\label{fig:MulCell_Coop_ISAC_Outage}
\end{figure}
\par
In Fig. \ref{fig:MulCell_Coop_ISAC_Outage}, we investigate the infeasibility rate of the optimization problem \eqref{prob1} versus the maximum tolerable interference for echo signal detection for different schemes. In particular, for different values of $I_{\mathrm{S}}^{\mathrm{tol}}$, we generate 100 random channel realizations. Then, we solve \eqref{prob1} and count the number of infeasible solutions.\footnote{For baseline scheme 2, we tackle \eqref{prob1} for all three BSs and count it as feasible as long as one of the three solutions is feasible.} As can be observed from the figure, compared to the two baseline schemes, the proposed scheme can significantly reduce the infeasibility rate thanks to the proposed multistatic architecture. As for the two baseline schemes, although the ISAC signal-induced SI is significantly suppressed by a factor of $100$ dB, the residual SI and unfavorable clutter can still severely impair the detection of the echo signal which suffers from the round-trip path loss. As a result, in the considered range of $I_{\mathrm{S}}^{\mathrm{tol}}$, both the baseline schemes rarely produce a feasible solution.
\section{Conclusions}
In this paper, we proposed a coordinated multi-cell network-based multistatic architecture to realize high-quality ISAC. In particular, we first developed a distance-based BS selection criterion to form a multistatic radar system. An AO-based computationally-efficient algorithm was developed to obtain a suboptimal transmit and receive beamforming policy which minimized the beam pattern mismatch error while satisfying ISAC-oriented requirements. Simulation results confirmed the effectiveness of the proposed ISAC framework and the corresponding resource allocation algorithm. Moreover, our results revealed the fact that the commonly adopted monostatic architecture-based framework may not provide reliable ISAC services for practical wireless networks. 
\bibliographystyle{IEEEtran}
\bibliography{Reference_List}
\end{document}